\documentclass[
 twocolumn, 
 reprint,
 superscriptaddress,
 amsmath,amssymb,
 aps,
 pra, 
]{revtex4-2}

\usepackage{graphicx}
\usepackage{dcolumn}
\usepackage{multirow} 
\usepackage{xcolor}
\usepackage[normalem]{ulem}
\usepackage{bm}
\usepackage{hyperref}
\usepackage{physics}
\usepackage{braket}
\usepackage{amsmath}
\usepackage{amsfonts}
\usepackage{amssymb}
\usepackage{amsthm}
\usepackage{mathtools}
\usepackage{color} 
\usepackage{gensymb}
\usepackage{here}

\begin{document}

\title{Size-consistent implementation of Hamiltonian simulation-based quantum-selected configuration interaction method for the supramolecular approach}

\author{Kenji Sugisaki}
\email{kensugisaki@tohmatsu.co.jp}
\affiliation{Deloitte Tohmatsu LLC, 3-2-3 Marunouchi, Chiyoda-ku, Tokyo 100-8363, Japan}
\affiliation{Centre for Quantum Engineering, Research and Education (CQuERE), TCG Centres for Research and Education in Science and Technology (TCG CREST), Sector V, Salt Lake, Kolkata 700091, India}

\date{\today}

\begin{abstract}

The quantum-selected configuration interaction (QSCI) method is a promising approach for large-scale quantum chemical calculations on currently available quantum hardware. However, its naive implementation lacks size consistency, which is essential for accurate intermolecular interaction energy calculations using the supramolecular approach. Here, we present a size-consistent implementation of QSCI by sampling Slater determinants for the dimer in the localized molecular orbital basis, constructing the subspaces for the monomers and dimer, and augmenting the dimer subspace with additional determinants required for size consistency. Implemented within the Hamiltonian simulation-based QSCI (HSB-QSCI) framework, our method numerically satisfies size consistency for 4H/8H/12H clusters, the FH dimer, and the FH--H$_2$O system. Application to intermolecular interaction energy calculations of hydrogen-bonded FH dimer and FH--H$_2$O demonstrates that our approach reproduces complete active space-configuration interaction (CAS-CI) values with errors below 0.04 kcal mol$^{-1}$.

\end{abstract}
\maketitle
\section{Introduction}

Quantum computation holds the potential to revolutionize our society. Among its diverse applications, sophisticated quantum chemical calculations are considered one of the most promising. Accurate quantum chemical calculations can accelerate drug and materials discovery, elucidate reaction mechanisms in enzymes and biomolecules, and so on. Quantum phase estimation (QPE)~\cite{Abrams-1999, Aspuru-Guzik-2005, Nielsen-Chuang} and variational quantum eigensolver (VQE)~\cite{Peruzzo-2014} are two major quantum computational approaches, targeting the fault-tolerant quantum computing (FTQC) and noisy intermediate-scale quantum (NISQ) eras, respectively, and both methods have been extensively investigated from theoretical and hardware demonstration perspectives. However, QPE typically requires quantum circuit that are prohibitively deep for even the most advanced quantum hardware available today, rendering its application to industrially relevant molecules extremely challenging until FTQC becomes a reality. On the other hand, VQE can be executed on NISQ devices, but it faces several significant challenges, such as substantial measurement overhead~\cite{Gonthier-2022}, the presence of barren plateaus~\cite{McClean-2018}, and so on. 

The quantum-selected configuration interaction (QSCI) method was proposed to perform large-scale quantum chemical calculations using currently available quantum hardware~\cite{Kanno-2023}. In QSCI, a quantum computer is used to sample Slater determinants that make significant contributions to the wave function of the target electronic state. The information from these sampled Slater determinants is then transferred to a classical computer, where the subspace Hamiltonian is constructed and diagonalized to obtain the energy and wave function. The key step in QSCI is the preparation of a quantum state suitable for sampling important Slater determinants. Initially, quantum state preparation methods based on VQE and adiabatic state preparation were proposed for this purpose~\cite{Kanno-2023}. In 2025, Robledo-Moreno and coworkers introduced a method for quantum state preparation based on the local unitary cluster Jastrow (LUCJ) ansatz~\cite{Motta-2023} with classically computed coupled cluster singles and doubles (CCSD) excitation amplitudes. They demonstrated QSCI calculations using up to 77 qubits on superconducting quantum processors, aided by an error mitigation technique called self-consistent configuration recovery~\cite{Robledo-Moreno-2025}. Recently, we proposed a quantum state preparation method for QSCI based on Hamiltonian simulation (HSB-QSCI)~\cite{Sugisaki-2025}, in which important Slater determinants are sampled from quantum states generated by real-time evolution of an approximate wave function. HSB-QSCI offers several advantages over the original QSCI: (1) a simple wave function such as a Hartree--Fock (HF) wave function can be used as the initial state for time evolution, (2) VQE variational optimization is not required, (3) different samples can be obtained by varying the evolution time, (4) long-time evolution is challenging to simulate on classical computers, and (5) higher-order excited configurations can be sampled even with shorter evolution times. We have also reported proof-of-principle demonstrations of HSB-QSCI for one-dimensional carbon chain molecules (carbynes), utilizing up to 36 qubits of an IBM superconducting quantum processor, with the aid of matrix product operator (MPO)-based classical optimization of the quantum circuit for the time evolution operator~\cite{Kanno-2025}. Shortly after our publication, methods based on the same concept were reported by other groups~\cite{Mikkelsen-2024, Yu-2025, Piccinelli-2025}. Notably, Weaving and coworkers reported that time evolution-based state preparation for QSCI yields a more compact wave function expansion than conventional selected CI in SiH$_4$ system~\cite{Weaving-2025}.

Although QSCI has been actively investigated from both theoretical and experimental perspectives~\cite{Nakagawa-2024, Liepuoniute-2025, Nutzel-2025, Kaliakin-2024, Barison-2025, Shajan-2025, Reinholdt-2025, Kaliakin-2025, Bauer-2025, Duriez-2025, Ohgoe-2025, Nogaki-2025, Bickley-2025, Chen-2025, Shivpuje-2025, Smith-2025, Matsuyama-2025}, it suffers from a serious drawback: QSCI is not size consistent unless special care is taken. In its naive implementation, the energy of a dimer composed of non-interacting monomers is generally not equal to the sum of the energies of the individual monomers. In fact, as discussed in the Supporting Information, the number of quantum circuit executions must increase exponentially with system size even approximately satisfy the size consistency condition. Consequently, the straightforward application of QSCI to the calculation of intermolecular interaction energies using the supramolecular approach is fundamentally impractical. In the supramolecular approach, the intermolecular interaction energy $E_\text{int}$ is calculated using the following equation: 
\begin{eqnarray}
    E_\text{int} = E_\text{AB} - E_\text{A} - E_\text{B}
    \label{eq:eq1}
\end{eqnarray}
Here, $E_\text{A}$, $E_\text{B}$, and $E_\text{AB}$ represent the energies of monomer A, monomer B, and the interacting dimer $\text{(A + B)}$, respectively. To address the challenges arising from the lack of size consistency in QSCI, the intermolecular interaction energy has instead been calculated using the following equation, 
\begin{eqnarray}
    E_\text{int} = E_\text{AB} - E_{\text{A}\cdots\text{B}}
    \label{eq:eq2}
\end{eqnarray}
Here, $E_{\text{A}\cdots\text{B}}$ denotes the energy of the dimer composed of monomers A and B at a sufficiently large inter-monomer distance. Hereafter, we refer to the approach based on equation (\ref{eq:eq2}) as the ``dimer approach''.
In 2025, Kaliakin and coworkers reported QSCI-based calculations of intermolecular interaction energies for water and methane dimers using the dimer approach~\cite{Kaliakin-2024}. However, it is evident that the computational cost is lower for the supramolecular approach than for the dimer approach, since the size of the subspace Hamiltonian to be diagonalized is smaller for monomers than for the dimer. It is also important to note that if Slater determinants are sampled independently for different geometries, the electronic configurations included in the subspace Hamiltonian may differ between geometries, potentially resulting in discontinuities in the potential energy curve. It should be emphasized that these issues also arise in other QSCI variants, as long as the selected CI framework is employed. In this work, we propose a size-consistent implementation of HSB-QSCI (sc-HSB-QSCI) to overcome these challenges. 

The rest of the paper is organized as follows. Section II provides a brief overview of quantum computational approaches to intermolecular interaction energy calculations reported to date, highlighting both their potential advantages and methodological challenges. In Section III, we discuss the theoretical foundations of the size-consistent implementation of HSB-QSCI. Section IV describes the computational conditions for proof-of-principle numerical simulations performed using a statevector simulator. Section V presents the results of these simulations, including validation of size consistency and application of the proposed method to intermolecular interaction energy calculations for hydrogen-bonded systems. Section VI concludes with a summary of our findings and outlines future prospects.

\section{Quantum computational approaches for intermolecular interaction energy calculations}

In this section, we present a brief overview of quantum computational approaches to intermolecular interaction energy calculations.

There are two major approaches to intermolecular interaction energy calculations, namely the symmetry-adapted perturbation theory (SAPT) and the supramolecular approach. SAPT is a perturbation theory-based method that evaluates intermolecular interaction energies using the wave functions of the individual monomers as the unperturbed states. Because SAPT defines the wave functions and energies for each monomer and introduces interaction terms perturbatively, the intermolecular interaction energy becomes zero when the molecules are infinitely separated, thereby ensuring size consistency automatically. Within the VQE framework, one- and two-electron reduced density matrices (1RDM and 2RDM, respectively) can be obtained via sampling on a quantum computer, making the application of VQE to SAPT relatively straightforward~\cite{Malone-2022, Loipersberger-2023}. In the context of QPE, Cortes and coworkers have reported an attempt to combine QPE with SAPT~\cite{Cortes-2024}. However, since SAPT requires the calculation of expectation values of the electrostatic ($\hat{V}$), exchange ($\hat{P}$), and electrostatic exchange ($\widehat{VP}_S$) operators, it must be integrated with expectation value estimation algorithms~\cite{Steudtner-2023}. This makes practical implementation on quantum hardware challenging, unless fault-tolerant quantum computing (FTQC) is realized. To the best of our knowledge, there are no reports of studies combining the QSCI method with SAPT.

In the supramolecular approach, the intermolecular interaction energy is calculated using equation (1), as discussed in the Introduction section. While this approach appears straightforward for evaluating intermolecular interaction energies, particular attention must be paid in quantum computations to ensure that the size consistency condition is satisfied. In fact, although the unitary coupled cluster (UCC) ansatz in VQE has generally been regarded as size-consistent due to its foundation in coupled cluster theory, our recent work~\cite{Sugisaki-2024a} revealed that size consistency in VQE-UCC can be broken when the Trotter decomposition is used to construct the quantum circuit. We found that using molecular orbitals localized on each monomer is essential for maintaining size consistency. In the case of QPE, when the quantum circuit for the time evolution operator is constructed via Trotter decomposition, not only the order of the Trotter decomposition (first or second order) and the number of Trotter slices, but also the choice of molecular orbitals for wave function expansion (RHF canonical orbitals or localized orbitals) and the term ordering in the Trotterized time evolution operator can impact size consistency condition~\cite{Sugisaki-2024b}. Intermolecular interaction energy calculations based on the supramolecular approach using QPE have been reported recently~\cite{Tachi-2025}. 

Thus, based on an overview of the field of quantum computational chemistry, the calculation of intermolecular interaction energies using quantum computers remains an area of active development. Consequently, the development of computational methods capable of accurately and efficiently evaluating intermolecular interaction energies on quantum computers represents a significant and ongoing challenge.

\section{Theory}

In conventional HSB-QSCI~\cite{Sugisaki-2025}, Slater determinants used for subspace Hamiltonian diagonalization are sampled from quantum states generated by real-time evolution of an approximate wave function. To expand the subspace, measurement outcomes obtained at  different evolution times, $k\Delta t$ for $k \le K$, are combined. For example, HSB-QSCI with $K = 3$ indicates that the subspace Hamiltonian is constructed using the measurement results from quantum circuit executions at $k$ = 1, 2, and 3. In the original proposal of the QSCI method~\cite{Kanno-2023}, only the $R$ most frequently observed Slater determinants are selected to define the subspace. In our HSB-QSCI implementation~\cite{Sugisaki-2025}, we do not impose such a threshold, and all sampled Slater determinants are included in the subspace.

Because the wave functions are simultaneous eigenfunctions of both the Hamiltonian and the electron spin ${\bf S}^2$ operator (hereafter referred to as spin eigenfunctions), whereas open-shell Slater determinants with spin-$\beta$ unpaired electrons are not. Without loss of generality, we assume $n_\alpha \ge n_\beta$, where $n_\alpha$ and $n_\beta$ denote the numbers of spin-$\alpha$ and spin-$\beta$ electrons, respectively. Due to the inherently stochastic nature of sampling on a quantum computer, it is possible that only a subset of the Slater determinants required to construct spin eigenfunctions will be sampled. To address this, HSB-QSCI introduces a symmetry completion step prior to constructing the subspace Hamiltonian on a classical computer. During this step, if the $|2\alpha\beta\alpha\beta0\rangle$ configuration (where 2, $\alpha$, $\beta$, and 0 indicate that the corresponding molecular orbital is doubly occupied, singly occupied by a spin-$\alpha$ electron, singly occupied by a spin-$\beta$ electron, and unoccupied, respectively) is sampled, then the configurations $|2\alpha\alpha\beta\beta0\rangle$, $|2\alpha\beta\beta\alpha0\rangle$, $|2\beta\alpha\alpha\beta0\rangle$, $|2\beta\beta\alpha\alpha0\rangle$, and $|2\beta\alpha\beta\alpha0\rangle$ are also added if they were not sampled. 

\begin{figure*}
    \centering
    \includegraphics[width=\linewidth]{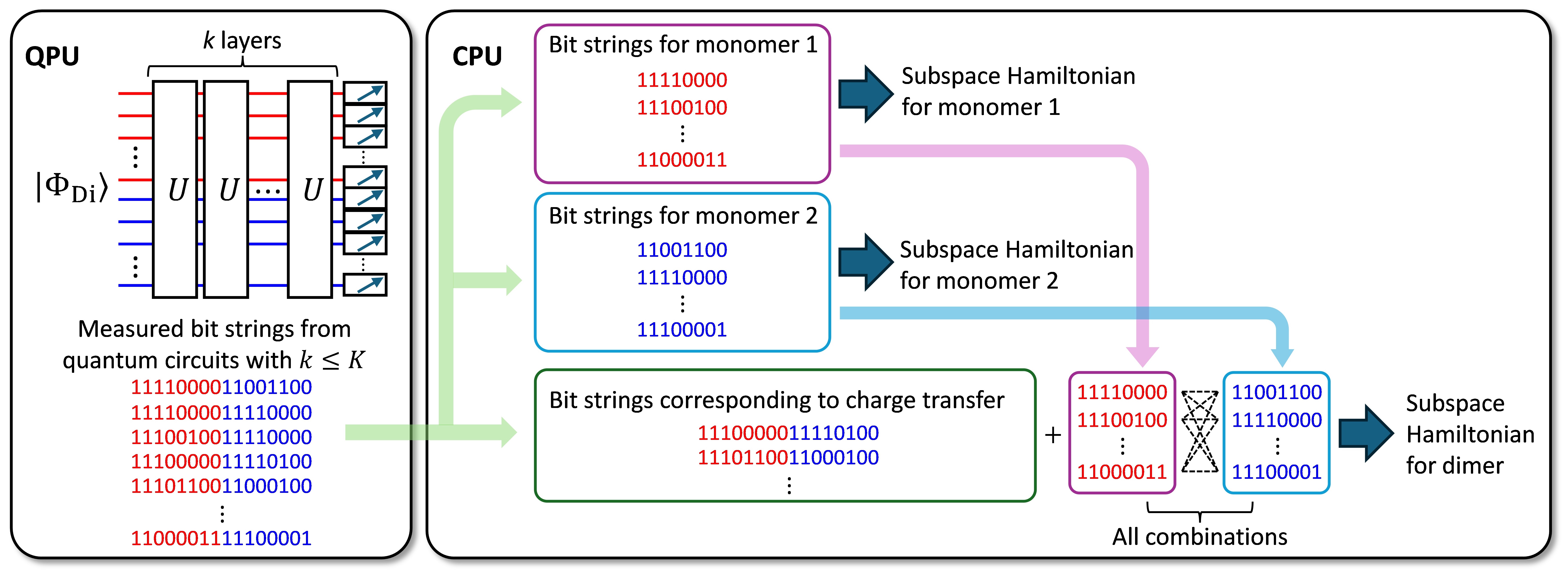}
    \caption{Schematic illustration of subspace generation in sc-HSB-QSCI. $|\Phi_\text{Di}\rangle$ represents an approximate wave function for the dimer, and $U$ is a time evolution operator $U = e^{-iHt}$.}
    \label{fig:fig1}
\end{figure*}

A schematic illustration of subspace construction in sc-HSB-QSCI is shown in Figure \ref{fig:fig1}. Note that the following subspace construction approach is not limited to HSB-QSCI and can be universally applied to any selected CI framework. The strategy of sc-HSB-QSCI is as follows: (1) Sample Slater determinants for the dimer in the localized molecular orbital basis. (2) For each sampled bit string, check the electronic configuration of the dimer. If the configuration involves only intra-monomer excitations from the initial state, split the bit string into those corresponding to each monomer and add them to the respective monomer subspace in the localized molecular orbital basis. If inter-monomer charge-transfer excitations are present, add the bit string to the dimer subspace without splitting. Note that when the Jordan--Wigner transformation is used for fermion--qubit encoding, the Hamming weight corresponds to the number of electrons. As a result, it becomes straightforward to assign electronic configurations to intra-monomer or inter-monomer excitations. (3) Perform subspace Hamiltonian diagonalization for the monomers using the Slater determinants generated in the previous step. (4) The dimer subspace Hamiltonian is constructed by including all possible combinations of the electronic configurations from monomer 1 and monomer 2 subspaces, as well as the Slater determinants corresponding to charge-transfer excitations. With this approach, electronic configurations required for size consistency but not sampled are systematically added, thereby ensuring that size consistency is automatically satisfied. 

In sc-HSB-QSCI, it is assumed that the molecular orbitals of the monomer and dimer are nearly identical. To satisfy this condition, we applied molecular orbital localization to both the monomers and the dimer in step 1. However, when two monomers are in close proximity and the inter-monomer interaction is significant, as in the hydrogen-bonded systems studied here, the localized orbitals cannot be completely confined to each monomer and instead exhibit small amplitudes on the other monomer, which may affect size consistency. To improve the locality of the molecular orbitals on each monomer, introducing a penalty term to the cost function when molecular orbitals are delocalized over both monomers could be a viable approach, which remains a topic for future investigation.

\section{Computational conditions}

As a proof-of-principle demonstration of sc-HSB-QSCI, we performed numerical simulations on a classical computer. The target systems included a dimer and a trimer of square tetrahydrogen (4H) clusters, hydrogen-bonded systems (FH dimer and FH--H$_2$O). The H--H bond length in the 4H cluster was set to 2.0 Bohr ($\approx$ 1.0583 \AA), and the dimer (8H cluster) and trimer (12H cluster) were constructed by placing two and three 4H clusters in a cuboid arrangement with an inter-monomer distance of 100 \AA. Note that 4H/8H clusters have previously been used to investigate size consistency in VQE with a unitary coupled cluster singles and doubles (UCCSD) ansatz~\cite{Sugisaki-2024a}, and in QPE~\cite{Sugisaki-2024b}. The localized molecular orbitals of the 8H and 12H clusters were manually generated by taking appropriate linear combinations of RHF/STO-3G~\cite{STO-3G} canonical orbitals. For sc-HSB-QSCI, all molecular orbitals were included in the Hamiltonian simulation, and the interaction energies were compared with those from full-configuration interaction (full-CI) calculations. 

\begin{figure}
    \centering
    \includegraphics[width=\linewidth]{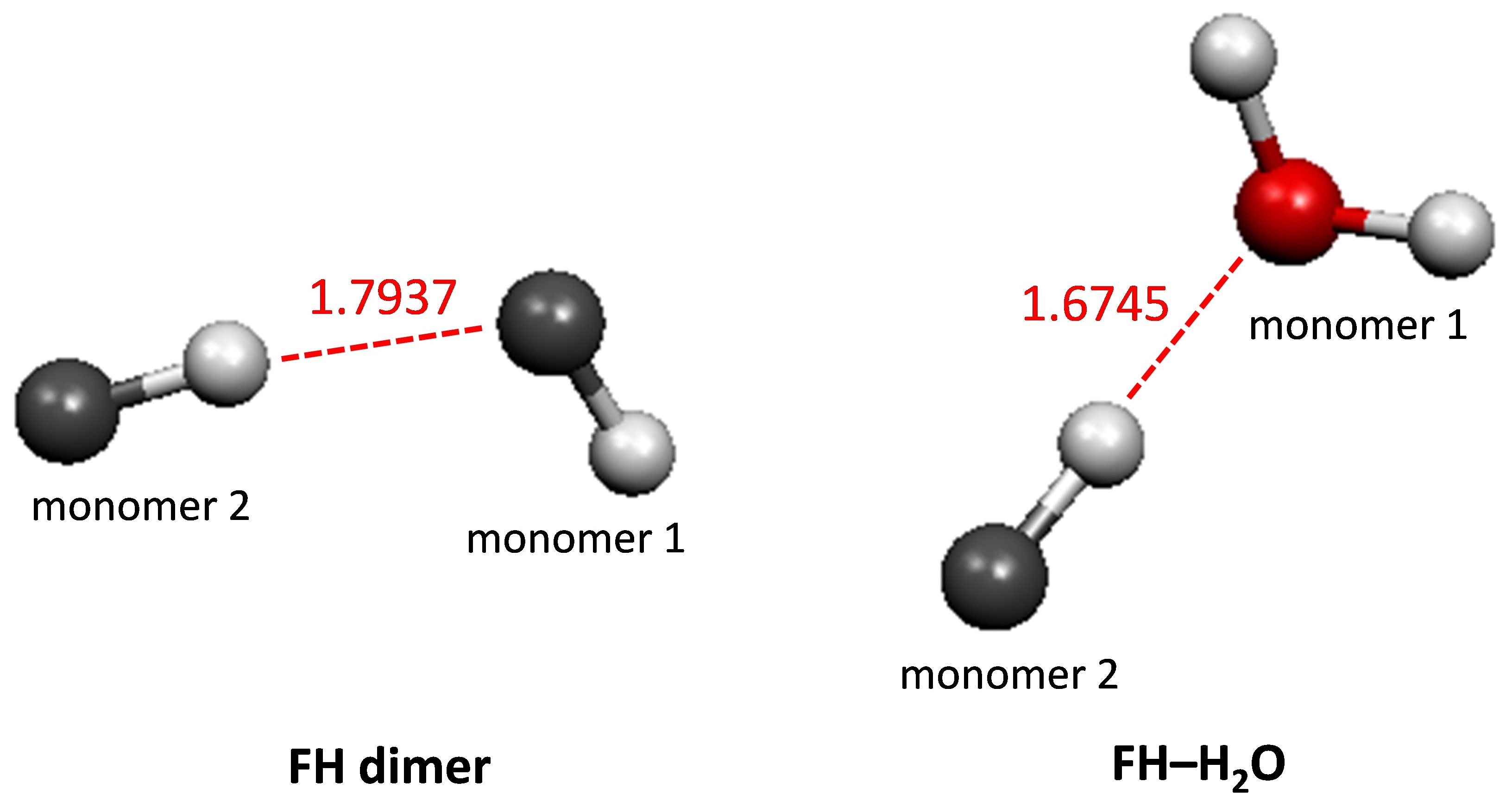}
    \caption{PBE0/aug-cc-pVDZ optimized geometry of the FH dimer and FH--H$_2$O. Red dotted lines represent hydrogen bonds, and values in red denote intermolecular distances in units of \AA.}
    \label{fig:fig2}
\end{figure}

For the hydrogen-bonded systems, geometry optimizations of the dimers were performed at the PBE0~\cite{PBE0}/aug-cc-pVDZ~\cite{cc-pVDZ, aug_for_cc-pVDZ} level of theory. No imaginary vibrational frequencies were obtained at the optimized structure. The optimized geometries are shown in Figure~\ref{fig:fig2}, and the corresponding Cartesian coordinates are provided in Tables \ref{tab:table_s1} and \ref{tab:table_s2} in the Supporting Information. For monomer calculations, we used the optimized geometry for the dimer without further performing geometry optimization of the monomers. 
To assess size consistency, we also generated geometries for the dimers without inter-monomer interaction by increasing the inter-monomer distance to 100 \AA. Localized molecular orbitals for both dimers and monomers were generated using the Pipek--Mezey method~\cite{Pipek-1989} at the RHF/aug-cc-pVDZ level. 
In the sc-HSB-QSCI simulations, only valence orbitals were considered, with active space sizes of (16e, 10o) for the FH dimer, (8e, 5o) for the FH monomer, (16e, 11o) for the FH--H$_2$O, and (8e, 6o) for the H$_2$O. Here, ($N$e, $M$o) indicates an active space consisting of $N$ electrons and $M$ molecular orbitals. The active orbitals for the monomers and dimers of the hydrogen-bonded FH dimer and FH--H$_2$O are shown in Figure~\ref{fig:fig_s2} in Supporting Information. All DFT, RHF, and MP2 calculations were performed using the GAMESS-US program package~\cite{GAMESS}. 

Numerical simulations for sc-HSB-QSCI were performed using our in-house Python3 code, developed with the OpenFermion~\cite{OpenFermion}, Cirq~\cite{Cirq}, and PyCI~\cite{PyCI} libraries. Details of the HSB-QSCI implementation can be found in Reference~\cite{Sugisaki-2025}. The Jordan--Wigner transformation~\cite{Jordan-1928} was used for fermion--qubit encoding. The RHF wave function served as the initial state for time evolution, and the quantum circuit for the time evolution operator $U = e^{-iH\Delta t}$ with $\Delta t = 1$ was constructed using the second-order Trotter--Suzuki decomposition with a single Trotter step. Unless otherwise noted, the number of shots for quantum circuit execution was set to $1 \times 10^4$, and $K$ is varied from 1 to 5 and 10. In sc-HSB-QSCI, symmetry completion~\cite{Sugisaki-2025} was performed prior to constructing the subspace Hamiltonian matrix, as discussed in the previous section.  

For the hydrogen-bonded FH dimer and FH--H$_2$O systems, we also calculated intermolecular interaction energies using heat-bath configuration interaction (HCI) in conjunction with the propose approach. The HCI calculations were performed using the PyCI~\cite{PyCI} library. 

\section{Results and Discussion}
\subsection{Size consistency of sc-HSB-QSCI}

\begin{table*}
\caption{\label{tab:table1} Sc-HSB-QSCI, RHF, and full-CI total energies, errors in the total energy with respect to the full-CI value ($\Delta E_{\text{sc-HSB-QSCI} - \text{full-CI}}$/Hartree) for the 8H (dimer) and 4H (monomer) clusters, in units of Hartree, and intermolecular interaction energy ($E_{int}$) in units of kcal mol$^{-1}$. The numbers in parenthesis of the total energy section indicate the number of Slater determinants included in the subspace Hamiltonian diagonalization.}
\begin{tabular*}{\textwidth}{@{\extracolsep{\fill}}clllcccc}
\hline
\multirow{2}{*}{$K$} & \multicolumn{3}{c}{Total energy/Hartree}               & \multicolumn{3}{c}{$\Delta E_{\text{sc-HSB-QSCI} - \text{full-CI}}$/Hartree}& \multirow{2}{*}{$E_{int}$/kcal mol$^{-1}$} \\
        & Dimer             &  Monomer 1       & Monomer 2        & Dimer    & Monomer 1 & Monomer 2 &   \\
 \hline
  1     & $-$3.652202 (98)  & $-$1.816720 (8)  & $-$1.835483 (8)  & 0.226661 & 0.122712  & 0.103949  & 0.0000 \\
  2     & $-$3.862635 (124) & $-$1.933870 (9)  & $-$1.928764 (10) & 0.016229 & 0.005561  & 0.010667  & 0.0000 \\
  3     & $-$3.862635 (124) & $-$1.933870 (9)  & $-$1.928764 (10) & 0.016229 & 0.005561  & 0.010667  & 0.0000 \\
  4     & $-$3.862664 (134) & $-$1.933900 (10) & $-$1.928764 (10) & 0.016199 & 0.005532  & 0.010667  & 0.0000 \\
  5     & $-$3.862664 (134) & $-$1.933900 (10) & $-$1.928764 (10) & 0.016199 & 0.005532  & 0.010667  & 0.0000 \\
\hline
 RHF    & $-$3.553541 (1)   & $-$1.776770 (1)  & $-$1.776770 (1)  &   &    &    & 0.0000 \\
full-CI & $-$3.878863 (178) & $-$1.939432 (12) & $-$1.939432 (12) &   &    &    & 0.0000 \\ 
\hline
\end{tabular*}
\end{table*}

Here, we demonstrate that the sc-HSB-QSCI method satisfies the size consistency condition, in the 8H cluster using the 4H cluster as the monomer. The square 4H cluster is a well-known strongly correlated system~\cite{Paldus-1993}, with a ground-state wave function exhibiting significant multi-configurational character. Using the STO-3G basis set~\cite{STO-3G}, the full-CI dimensions are 12 and 178 for the monomer (4H cluster in the $D_{2h}$ point group) and the dimer (8H cluster in the $C_{2v}$ point group), respectively. Owing to the small system size, sc-HSB-QSCI calculations were performed with 10 shots for each $k$. The results are summarized in Table \ref{tab:table1}. The computed $E_\text{int}$ values are less than $1 \times 10^{-7}$ Hartree for all $K$. The small deviations of $E_\text{int}$ from zero arize from rounding errors in the one- and two-electron integrals. These results clearly demonstrate the size consistent nature of sc-HSB-QSCI. 
Note that the 8H cluster consists of two identical monomers (4H clusters), yet the sc-HSB-QSCI energies in Table \ref{tab:table1} do not yield identical energies for the two monomers. This discrepancy arises because the sampling process in sc-HSB-QSCI is inherently stochastic, and symmetrically equivalent molecules can exhibit different energies if the sets of sampled Slater determinants differ. Although point group symmetry is not considered in this work, it is possible to leverage molecular point group symmetry to ensure that two equivalent molecules have the same energies. 

Importantly, the present implementation consistently satisfies the size consistency condition, regardless of the system size or the number of monomers, because the electronic configurations required for size consistency are automatically included in the subspace. To demonstrate this, we performed numerical simulations for the 12H cluster as the trimer and evaluated the size consistency error by calculating the following quantities. 
\begin{eqnarray}
    \Delta E_{123} = E(123) - E(1) - E(2) - E(3) 
\end{eqnarray}
\begin{eqnarray}
    \Delta E_{12} = E(12) - E(1) - E(2)
\end{eqnarray}
\begin{eqnarray}
    \Delta E_{13} = E(13) - E(1) - E(3)
\end{eqnarray}
\begin{eqnarray}
    \Delta E_{23} = E(23) - E(2) - E(3)
\end{eqnarray}
Here, $E(1)$, $E(2)$, and $E(3)$ denote the total energies of monomers 1, 2, and 3, respectively, $E(12)$, $E(13)$, and $E(23)$ represent the energies of the dimers composed of monomers 1 and 2, 1 and 3, and 2 and 3, respectively, and $E(123)$ is the energy of the trimer. Table \ref{tab:table2} summarizes the simulation results, which yield zero intermolecular interaction energies for all systems.

\begin{table}
\caption{\label{tab:table2} Intermolecular interaction energies for the 12H cluster consisting of three 4H clusters at inter-monomer distances of 100 \AA\, in units of kcal mol$^{-1}$.}
\begin{tabular*}{\linewidth}{@{\extracolsep{\fill}}crrrr}
\hline
$K$ & $\Delta E_{123}$ & $\Delta E_{12}$ & $\Delta E_{13}$ & $\Delta E_{23} $\\
\hline
1  & 0.0000 & 0.0000 & 0.0000 & 0.0000 \\
2  & 0.0000 & 0.0000 & 0.0000 & 0.0000 \\
3  & 0.0000 & 0.0000 & 0.0000 & 0.0000 \\
4  & 0.0000 & 0.0000 & 0.0000 & 0.0000 \\
5  & 0.0000 & 0.0000 & 0.0000 & 0.0000 \\
10 & 0.0000 & 0.0000 & 0.0000 & 0.0000 \\
\hline
\end{tabular*}
\end{table}

\begin{table}
\caption{\label{tab:table3} Intermolecular interaction energies for the FH dimer and FH--H$_2$O at an inter-monomer distance of 100 \AA, calculated using sc-HSB-QSCI and org-HSB-QSCI, in units of kcal mol$^{-1}$.}
\begin{tabular*}{\linewidth}{@{\extracolsep{\fill}}ccrr}
\hline
Molecule & $K$ & sc-HSB-QSCI & org-HSB-QSCI \\
\hline
FH dimer & 1 & 0.000 & 0.016 \\
         & 2 & 0.000 & 0.016 \\
         & 3 & 0.000 & 0.016 \\
         & 4 & 0.000 & 0.013 \\
         & 5 & 0.000 & 0.007 \\
FH--H$_2$O & 1 & 0.000 & 0.095 \\
           & 2 & 0.000 & 0.046 \\
           & 3 & 0.000 & 0.040 \\
           & 4 & 0.000 & 0.040 \\
           & 5 & 0.000 & 0.031 \\
\hline
\end{tabular*}
\end{table}

Next, we numerically investigated size consistency in the FH dimer and FH--H$_2$O systems using both sc-HSB-QSCI and the original form of HSB-QSCI (org-HSB-QSCI). As discussed above, org-HSB-QSCI is not size consistent, and the interaction energy can be nonzero even when the intermolecular distance is sufficiently large. In this study, we used geometries with an intermolecular distance of 100 \AA. The calculated interaction energies are summarized in Table \ref{tab:table3}, and total energies and the number of Slater determinants in the subspace are provided in Tables \ref{tab:table_s3} and \ref{tab:table_s4} in the Supporting Information. Sc-HSB-QSCI yielded an interaction energy $E_\text{int}$ = 0.000 kcal mol$^{-1}$ for all $K$, whereas org-HSB-QSCI gave nonzero interaction energies. The error in $E_\text{int}$ is slightly larger for FH--H$_2$O than for the FH dimer, suggesting that the violation of size consistency can increase as the number of active orbitals grows. It should also be noted that the energy error arising from violation of size consistency in org-HSB-QSCI is less than 0.1 kcal mol$^{-1}$ in Table \ref{tab:table3}, but this is because a relatively large number of shots was used. Table \ref{tab:table4} summarizes the dependence of the intermolecular interaction energies on the number of shots for the FH dimer and FH--H$_2$O at an inter-monomer distance of 100 \AA. It is clear that violation of size consistency become more pronounced when a smaller number of shots is employed. There results also suggest that violation of size consistency becomes more significant as the active space increases. In all systems under study, the number of sampled charge-transfer excited Slater determinants was zero. 

\begin{table}
\caption{\label{tab:table4} Shot number dependence on intermolecular interaction energies for the FH dimer and FH--H$_2$O at an inter-monomer distance of 100 \AA, calculated using sc-HSB-QSCI and org-HSB-QSCI at $K = 5$, in units of kcal mol$^{-1}$.}
\begin{tabular*}{\linewidth}{@{\extracolsep{\fill}}ccrr}
\hline
Molecule & \#shots & sc-HSB-QSCI & org-HSB-QSCI \\
\hline
FH dimer & 100   & 0.000 &    0.540 \\
         & 200   & 0.000 &    0.447 \\
         & 300   & 0.000 & $-$0.059 \\
         & 400   & 0.000 &    0.085 \\
         & 500   & 0.000 & $-$0.221 \\
         & 1000  & 0.000 &    0.231 \\
         & 10000 & 0.000 &    0.007 \\
FH--H$_2$O & 100   & 0.000 &    0.710 \\
           & 200   & 0.000 &    0.471 \\
           & 300   & 0.000 & $-$0.631 \\
           & 400   & 0.000 &    0.344 \\
           & 500   & 0.000 &    0.193 \\
           & 1000  & 0.000 & $-$0.143 \\
           & 10000 & 0.000 &    0.031 \\
\hline
\end{tabular*}
\end{table}

\subsection{Intermolecular interaction energies of hydrogen-bonded FH dimer and FH--H$_2$O}

Next, we examine the intermolecular interaction energy calculations of the hydrogen-bonded FH dimer and FH--H$_2$O. Hydrogen bonding is one of the most important intermolecular interactions, particularly in biomolecules. Here, we compare interaction energies calculated using the supramolecular approach with sc-HSB-QSCI and org-HSB-QSCI, as well as the dimer approach with org-HSB-QSCI. The numerical simulation results, together with RHF and CAS-CI interaction energies, are summarized in Table \ref{tab:table5}, and the total energies are listed in Tables \ref{tab:table_s5} and \ref{tab:table_s6} in the Supporting Information. Sc-HSB-QSCI yields larger interaction energies in absolute value and smaller errors relative to the CAS-CI values compared to org-HSB-QSCI. Org-HSB-QSCI using the dimer approach gave interaction energies comparable to those obtained with sc-HSB-QSCI, although its computational cost is higher than that of the supramolecular approach. At $K = 10$, sc-HSB-QSCI gave interaction energies with deviations of 0.012 and 0.038 kcal mol$^{-1}$ from the CAS-CI values for the FH dimer and FH--H$_2$O, respectively. The number of Slater determinants included in the subspace Hamiltonian diagonalization for the dimer is 458 and 4150 for the FH dimer and FH--H$_2$O, respectively, which should be compared to 2025 and 27225, respectively, at the CAS-CI level of theory. 

Table \ref{tab:table_s7} in the Supporting Information presents the classification of bit strings sampled by the quantum computer into those corresponding to intra-monomer excited determinants and those representing inter-monomer charge-transfer excited determinants.
Whether intra-monomer excitation or inter-monomer charge-transfer excitation is sampled more frequently depends on the duration of the time evolution. As $k$ increases, the probability of sampling charge-transfer excitation becomes higher. This indicates that longer time evolution is important for accurately obtaining intermolecular interaction energies using sc-HSB-QSCI.   

In the hydrogen-bonded FH dimer and FH--H$_2$O systems, one unoccupied molecular orbital is delocalized over the dimer (See the 10th and 11th LMOs of the FH dimer and FH--H$_2$O, respectively, in Figure S2 in the Supporting Information). The overlaps between the molecular orbitals of the monomer and dimer, $|\langle\varphi_{i;\text{Monomer}}|\varphi_{j;\text{Dimer}}\rangle|$, are calculated to be 0.959 and 0.840 for the FH dimer and FH--H$_2$O, respectively. Such molecular orbital delocalization indicates that the monomer and dimer do not share the same Hilbert space as the active space, which can influence the calculated intermolecular interaction energy. Generating nearly identical LMOs for both the monomer and dimer when the two molecules are in close proximity remains an unresolved issue. However, as discussed in the Theory section, this problem may be mitigated by designing a cost function for molecular orbital localization that includes a penalty when molecular orbitals are delocalized over the dimer. In contrast, in the absence of intermolecular interactions, such orbital delocalization does not occur, and the intermolecular interaction energy becomes zero.

\begin{table*}
\caption{\label{tab:table5} Intermolecular interaction energies of the hydrogen-bonded FH dimer and FH--H$_2$O, calculated using the supramolecular approach with sc-HSB-QSCI and org-HSB-QSCI, as well as the dimer approach with org-HSB-QSCI, in units of kcal mol$^{-1}$.}
\begin{tabular*}{\linewidth}{@{\extracolsep{\fill}}ccccccc}
\hline
\multirow{2}{*}{System} & \multirow{2}{*}{$K$} & \multicolumn{2}{c}{Supramolecular approach} & Dimer approach & \multirow{2}{*}{RHF} & \multirow{2}{*}{CAS-CI} \\
 & & sc-HSB-QSCI & org-HSB-QSCI & org-HSB-QSCI & & \\
\hline
 FH dimer  &  1 & $-4.030$ & $-4.002$ & $-4.019$ & \multirow{6}{*}{$-3.971$} & \multirow{6}{*}{$-4.071$} \\
           &  2 & $-4.037$ & $-4.011$ & $-4.027$ & & \\
           &  3 & $-4.043$ & $-4.020$ & $-4.036$ & & \\
           &  4 & $-4.055$ & $-4.040$ & $-4.054$ & & \\
           &  5 & $-4.056$ & $-4.041$ & $-4.048$ & & \\
           & 10 & $-4.059$ & $-4.049$ & $-4.053$ & & \\
FH--H$_2$O &  1 & $-8.450$ & $-8.140$ & $-8.234$ & \multirow{6}{*}{$-7.902$} & \multirow{6}{*}{$-8.786$} \\
           &  2 & $-8.497$ & $-8.402$ & $-8.447$ & & \\
           &  3 & $-8.687$ & $-8.600$ & $-8.640$ & & \\
           &  4 & $-8.695$ & $-8.617$ & $-8.658$ & & \\
           &  5 & $-8.717$ & $-8.674$ & $-8.705$ & & \\
           & 10 & $-8.748$ & $-8.717$ & $-8.741$ & & \\
\hline
\end{tabular*}
\end{table*}

Since our size-consistent implementation is applicable to other selected CI frameworks, we applied our proposed scheme to HCI~\cite{Holmes-2016}. Details of the numerical simulations are provided in subsection C of the Supporting Information. The size consistent HCI (sc-HCI) yielded intermolecular interaction energies slightly closer to the CAS-CI values than those obtained with sc-HSB-QSCI, indicating that sc-HCI achieves a more compact wave function expansion for a given level of accuracy. As discussed in our previous work~\cite{Sugisaki-2025}, HSB-QSCI may sample Slater determinants that are important for excited states rather than the ground state. Therefore, screening out unimportant Slater determinants from the subspace is essential for obtaining a more compact wave function expansion in sc-HSB-QSCI, which remains one of the most significant challenges in QSCI. Nevertheless, the approach proposed in this study significantly improves the applicability of selected CI methods to intermolecular interaction energy calculations.

\section{Conclusion}
In summary, we have developed a size-consistent implementation of Hamiltonian simulation-based quantum-selected configuration interaction (sc-HSB-QSCI) for intermolecular interaction energy calculations based on the supramolecular approach. The size consistency of the proposed method was confirmed through numerical simulations of the 8H and 12H clusters, FH dimer, and FH--H$_2$O systems with an inter-monomer distance of 100 \AA, all yielding zero intermolecular interaction energies. Interaction energies of intermolecular hydrogen bonds in the FH dimer and FH--H$_2$O were also calculated with sc-HSB-QSCI, and the results agreed with the CAS-CI interaction energies in the same active space with errors of only 0.012 and 0.038 kcal mol$^{-1}$ for the FH dimer and FH--H$_2$O, respectively. During the sampling procedure, intra-monomer excited configurations are predominantly sampled in the early stage of real-time evolution, while the probability of sampling charge-transfer excited configurations increases at later stages. This trend indicates that extended real-time evolution is essential for accurately computing intermolecular interaction energies with sc-HSB-QSCI. 

Apart from the supramolecular approach, size consistency is also crucial in other contexts, such as the fragment molecular orbital method~\cite{Kitaura-1999, Mochizuki-2022} and the ONIOM (our own N-layer integrated molecular orbitals and molecular mechanics) method~\cite{Svensson-1996, Chung-2015}. Applying the sc-HSB-QSCI method to these problems remains an important avenue for future research.

\section{Acknowledgments}
K.S. thanks Dr. Yuhei Tachi and Dr. Hiroyuki Tezuka for useful discussions.

\bibliography{ref}
\bibliographystyle{unsrt}

\clearpage

\section{Supporting Information}

\renewcommand{\thepage}{S\arabic{page}}
\setcounter{page}{1}
\renewcommand{\thefigure}{S\arabic{figure}}
\setcounter{figure}{0}
\renewcommand{\thetable}{S\arabic{table}}
\setcounter{table}{0}

\subsection{Cartesian coordinates}

\begin{table}[H]
\centering
    \caption{Cartesian coordinates of the PBE0/aug-cc-pVDZ optimized geometry for the hydrogen-bonded FH dimer, given in units of \AA.}
    \begin{tabular*}{\linewidth}{@{\extracolsep{\fill}}crrr}
    \hline
    Atom & \multicolumn{1}{c}{X} & \multicolumn{1}{c}{Y} & \multicolumn{1}{c}{Z} \\
    \hline
    F & $-$0.355909 &    0.404037 &    0.199053 \\
    H &    0.112061 & $-$0.309544 & $-$0.156637 \\
    F & $-$3.008949 & $-$0.078707 & $-$0.096402 \\
    H & $-$2.130203 &    0.184214 &    0.053986 \\
    \hline
    \end{tabular*}
    \label{tab:table_s1}
\end{table}

\begin{table}[H]
\centering
    \caption{Cartesian coordinates of the PBE0/aug-cc-pVDZ optimized geometry for the hydrogen-bonded FH--H$_2$O, given in units of \AA.}
    \begin{tabular*}{\linewidth}{@{\extracolsep{\fill}}crrr}
    \hline
    Atom & \multicolumn{1}{c}{X} & \multicolumn{1}{c}{Y} & \multicolumn{1}{c}{Z} \\
    \hline
    O & $-$2.451734 &    0.852831 & $-$0.885154 \\
    H & $-$1.585104 &    0.709253 & $-$0.491503 \\
    H & $-$2.774601 &    1.674026 & $-$0.500774 \\
    F & $-$4.071275 & $-$1.154030 & $-$0.438062 \\
    H & $-$3.491540 & $-$0.434027 & $-$0.626687 \\
    \hline
    \end{tabular*}
    \label{tab:table_s2}
\end{table}

\subsection{Relationship between the number of quantum circuit repetitions and size consistency}
Here, we demonstrate that the number of quantum circuit repetitions must increase exponentially with system size in order to approximately satisfy the size consistency condition in QSCI. For clarity, our discussion focuses on the original QSCI and assumes access to a state preparation oracle capable of generating the CAS-CI wave function. 
\begin{eqnarray}
    Prep|0\rangle^{\otimes n} = |\Psi\rangle = \sum_j c_j |\psi_j\rangle
    \notag
\end{eqnarray}
Here, $|\Psi\rangle$ denotes the CAS-CI wave function of the target electronic state, and $\{|\psi_j\rangle\}$ represents the set of Slater determinants. In addition, we assume an ideal measurement scenario in which each Slater determinant $|\psi_j\rangle$ is sampled once for every $1/|c_j|^2$ measurements performed. 

Let us consider a dimer composed of two monomers with no intermolecular interactions. In this case, the wave function of the dimer can be expressed as the direct product of the wave functions of the individual monomers, as shown below.
\begin{eqnarray}
    |\Psi_\text{dimer}\rangle 
    = \sum_{jk} c_j c_k |\psi_j\rangle \otimes |\psi_k\rangle
    \notag
\end{eqnarray}
Based on this, sampling the Slater determinant represented as the direct product $|\psi_j\rangle \otimes |\psi_j\rangle$ requires executing the quantum circuit $1/|c_j|^4$ times.

\subsection{Size-consistent implementation for heat-bath CI}

\begin{figure}[ht]
    \centering
    \includegraphics[width=\linewidth]{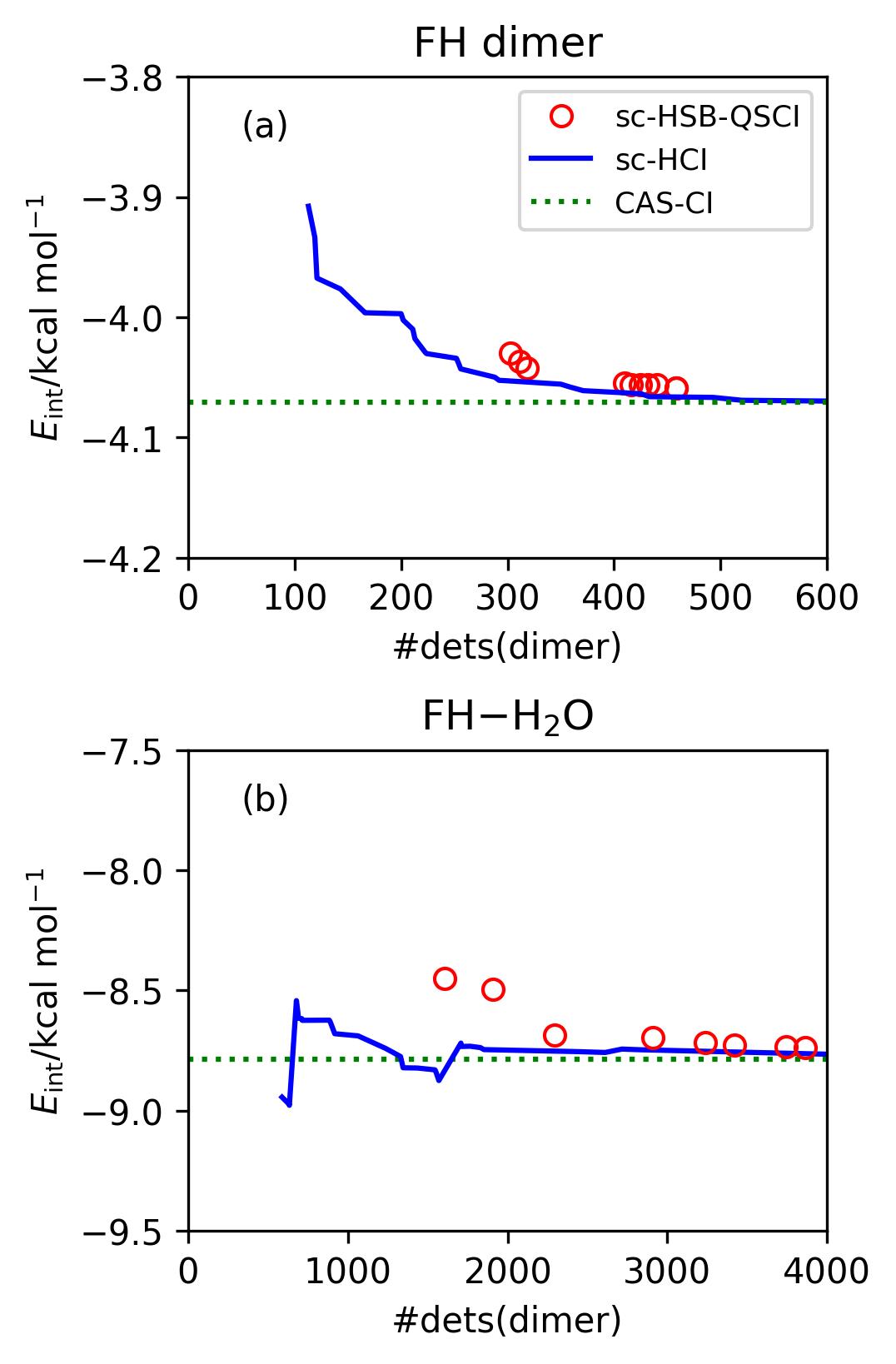}
    \caption{Number of dimer Slater determinants included in the subspace Hamiltonian versus intermolecular interaction energy obtained from sc-HCI and sc-HSB-QSCI in (a) FH dimer and in (b) FH--H$_2$O. }
    \label{fig:fig_s1}
\end{figure}

As discussed in the Theory section, the size-consistent implementation proposed in this study is applicable to any selected CI framework. Here, we examine its application to heat-bath CI (sc-HCI). In sc-HCI, we first perform HCI calculations for the dimer, and the resulting Slater determinants are used to construct the subspaces for both the monomers and the dimer. In this work, HCI was carried out with threshold values $\varepsilon$ ranging from $1.0 \times 10^{-4}$ to $5.0 \times 10^{-3}$. Figure~\ref{fig:fig_s1} plots the number of dimer Slater determinants against the intermolecular interaction energies calculated using both sc-HCI as well as sc-HSB-QSCI. Notably, sc-HCI yielded intermolecular interaction energies that are closer to the CAS-CI values than those obtained with sc-HSB-QSCI.  

%\subsection{Localized molecular orbitals and sc-HSB-QSCI results}

\begin{figure*}[ht]
    {\bf D.  Localized molecular orbitals and sc-HSB-QSCI results}
    \centering
    \includegraphics[width=\linewidth]{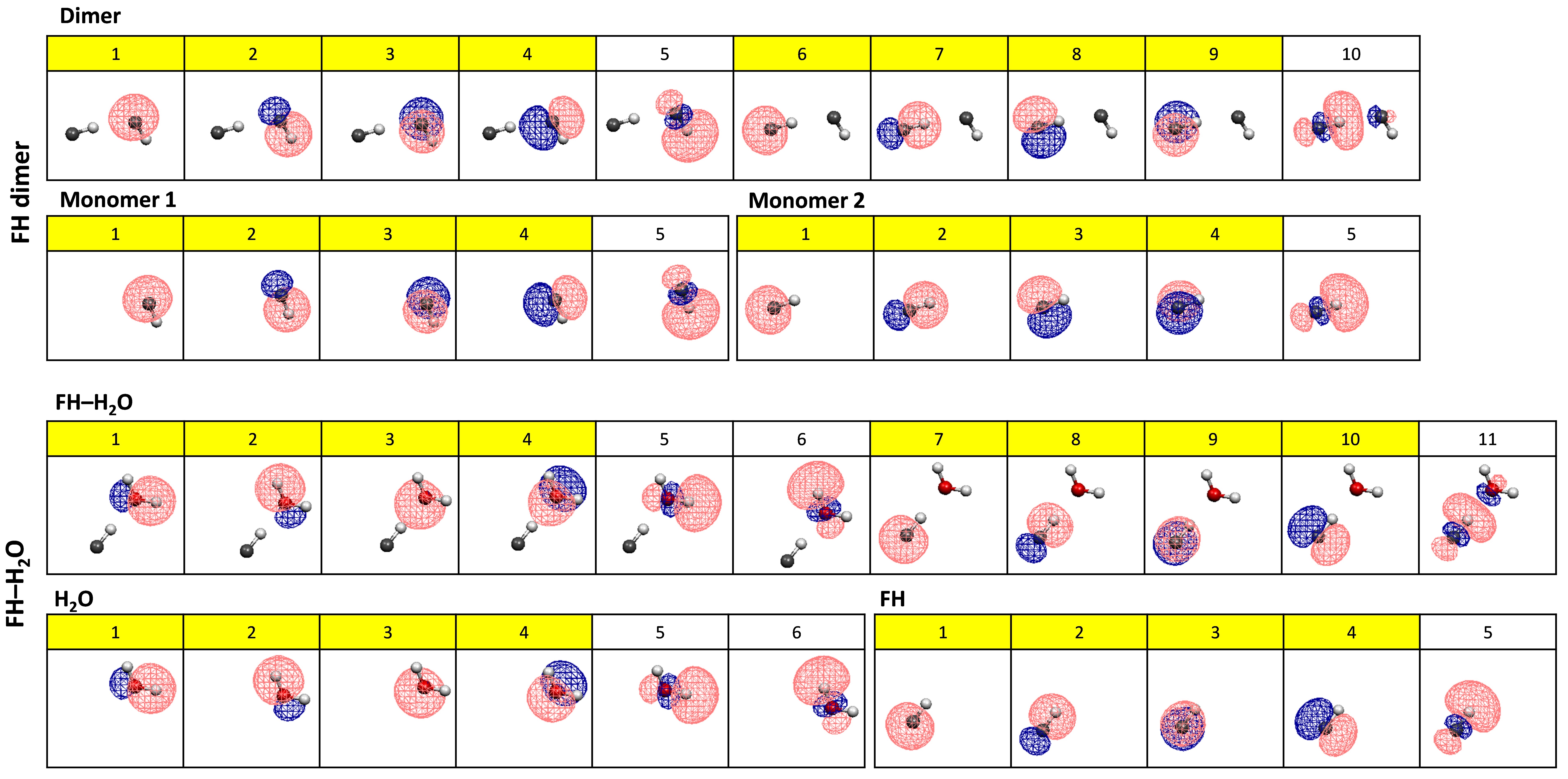}
    \caption{Localized molecular orbitals of FH dimer and FH--H$_2$O used as active space for HSB-QSCI. Orbitals with yellow highlights are occupied ones, and those without highlight are unoccupied. }
    \label{fig:fig_s2}
\end{figure*}

\begin{table*}[ht]
\caption{\label{tab:table_s3} Sc-HSB-QSCI, org-HSB-QSCI, RHF, and full-CI energies of the FH dimer at an intermolecular distance of 100 \AA, in units of Hartree. The numbers in parenthesis indicate the number of Slater determinants in the subspace.}
\begin{tabular*}{\linewidth}{@{\extracolsep{\fill}}cllllll}
\hline
\multirow{2}{*}{$K$} & \multicolumn{3}{c}{sc-HSB-QSCI} & \multicolumn{3}{c}{org-HSB-QSCI} \\
    & \multicolumn{1}{c}{$E$(dimer)} & \multicolumn{1}{c}{$E$(monomer 1)} & \multicolumn{1}{c}{$E$(monomer 2)} & \multicolumn{1}{c}{$E$(dimer)} & \multicolumn{1}{c}{$E$(monomer 1)} & \multicolumn{1}{c}{$E$(monomer 2)} \\
 \hline
  1 & $-200.097212$ (139) & $-100.048655$ (11) & $-100.048558$ (11) & $-200.097187$ (39) & $-100.048655$ (11) & $-100.048558$ (11) \\
  2 & $-200.097212$ (139) & $-100.048655$ (11) & $-100.048558$ (11) & $-200.097187$ (48) & $-100.048655$ (11) & $-100.048558$ (11) \\
  3 & $-200.097212$ (139) & $-100.048655$ (11) & $-100.048558$ (11) & $-200.097187$ (48) & $-100.048655$ (11) & $-100.048558$ (11) \\
  4 & $-200.097212$ (139) & $-100.048655$ (11) & $-100.048558$ (11) & $-200.097191$ (52) & $-100.048655$ (11) & $-100.048558$ (11) \\
  5 & $-200.097212$ (139) & $-100.048655$ (11) & $-100.048558$ (11) & $-200.097202$ (62) & $-100.048655$ (11) & $-100.048558$ (11) \\
\hline
RHF & $-200.065881$ (1) & $-100.033096$ (1) & $-100.032785$ (1) & $-200.065881$ (1) & $-100.033096$ (1) & $-100.032785$ (1) \\
full-CI & $-200.097212$ (2025) & $-100.048655$ (25) & $-100.048558$ (25) & $-200.097212$ (2025) & $-100.048655$ (25) & $-100.048558$ (25) \\
\hline
\end{tabular*}
\end{table*}

\begin{table*}[ht]
\caption{\label{tab:table_s4} Sc-HSB-QSCI, org-HSB-QSCI, RHF, and full-CI energies of the FH--H$_2$O at an intermolecular distance of 100 \AA, in units of Hartree. The numbers in parenthesis indicate the number of Slater determinants in the subspace.}
\begin{tabular*}{\linewidth}{@{\extracolsep{\fill}}cllllll}
\hline
\multirow{2}{*}{$K$} & \multicolumn{3}{c}{sc-HSB-QSCI} & \multicolumn{3}{c}{org-HSB-QSCI} \\
    & \multicolumn{1}{c}{$E$(dimer)} & \multicolumn{1}{c}{$E$(monomer 1)} & \multicolumn{1}{c}{$E$(monomer 2)} & \multicolumn{1}{c}{$E$(dimer)} & \multicolumn{1}{c}{$E$(monomer 1)} & \multicolumn{1}{c}{$E$(monomer 2)} \\
 \hline
  1 & $-176.128580$ (1327) & $-76.080556$ (95)  & $-100.048023$ (11) & $-176.128507$ (198) & $-76.080634$ (86)  & $-100.048023$ (11) \\
  2 & $-176.128687$ (1394) & $-76.080663$ (100) & $-100.048023$ (11) & $-176.128616$ (213) & $-76.080665$ (99)  & $-100.048023$ (11) \\
  3 & $-176.128697$ (1433) & $-76.080674$ (103) & $-100.048023$ (11) & $-176.128626$ (218) & $-76.080667$ (107) & $-100.048023$ (11) \\
  4 & $-176.128703$ (1545) & $-76.080680$ (111) & $-100.048023$ (11) & $-176.128641$ (246) & $-76.080683$ (119) & $-100.048023$ (11) \\
  5 & $-176.128710$ (1629) & $-76.080686$ (117) & $-100.048023$ (11) & $-176.128661$ (301) & $-76.080687$ (121) & $-100.048023$ (11) \\
\hline
RHF & $-176.072736$ (1) & $-76.041090$ (1) & $-100.031644$ (1) & $-176.072736$ (1) & $-76.041090$ (1) & $-100.031644$ (1) \\
full-CI & $-176.128718$ (27225) & $-76.080692$ (225) & $-100.048024$ (25) & $-176.128718$ (27225) & $-76.080692$ (225) & $-100.048024$ (25) \\
\hline
\end{tabular*}
\end{table*}

\begin{table*}[ht]
\caption{\label{tab:table_s5} Sc-HSB-QSCI, org-HSB-QSCI, RHF, and full-CI energies of the hydrogen-bonded FH dimer, in units of Hartree. The numbers in parenthesis indicate the number of Slater determinants in the subspace.}
\begin{tabular*}{\linewidth}{@{\extracolsep{\fill}}cllll}
\hline
\multirow{2}{*}{$K$} & \multicolumn{3}{c}{sc-HSB-QSCI} & \multicolumn{1}{c}{org-HSB-QSCI} \\
    & \multicolumn{1}{c}{$E$(dimer)} & \multicolumn{1}{c}{$E$(monomer 1)} & \multicolumn{1}{c}{$E$(monomer 2)} & \multicolumn{1}{c}{$E$(dimer)} \\
 \hline
  1 & $-200.103634$ (303) & $-100.048655$ (15) & $-100.048558$ (13) & $-200.103590$ (134) \\
  2 & $-200.103645$ (311) & $-100.048655$ (15) & $-100.048558$ (13) & $-200.103605$ (156) \\
  3 & $-200.103655$ (318) & $-100.048655$ (15) & $-100.048558$ (13) & $-200.103618$ (178) \\
  4 & $-200.103674$ (410) & $-100.048655$ (17) & $-100.048558$ (15) & $-200.103651$ (204) \\
  5 & $-200.103676$ (416) & $-100.048655$ (17) & $-100.048558$ (15) & $-200.103653$ (216) \\
 10 & $-200.103681$ (458) & $-100.048655$ (17) & $-100.048558$ (15) & $-200.103665$ (280) \\
\hline
RHF & $-200.072209$ (1) & $-100.033096$ (1) & $-100.032785$ (1) & $-200.072209$ (1) \\
full-CI & $-200.103699$ (2025) & $-100.048655$ (25) & $-100.048558$ (25) & $-200.103699$ (2025) \\
\hline
\end{tabular*}
\end{table*}

\begin{table*}[ht]
\caption{\label{tab:table_s6} Sc-HSB-QSCI, org-HSB-QSCI, RHF, and full-CI energies of the hydrogen-bonded FH--H$_2$O, in units of Hartree. The numbers in parenthesis indicate the number of Slater determinants in the subspace.}
\begin{tabular*}{\linewidth}{@{\extracolsep{\fill}}cllll}
\hline
\multirow{2}{*}{$K$} & \multicolumn{3}{c}{sc-HSB-QSCI} & \multicolumn{1}{c}{org-HSB-QSCI} \\
    & \multicolumn{1}{c}{$E$(dimer)} & \multicolumn{1}{c}{$E$(monomer 1)} & \multicolumn{1}{c}{$E$(monomer 2)} & \multicolumn{1}{c}{$E$(dimer)} \\
 \hline 
  1 & $-176.141994$ (1608) & $-76.080505$ (97)  & $-100.048023$ (11) & $-176.141629$ (514)  \\
  2 & $-176.142184$ (1911) & $-76.080621$ (112) & $-100.048023$ (11) & $-176.142078$ (685)  \\
  3 & $-176.142490$ (2293) & $-76.080624$ (132) & $-100.048023$ (11) & $-176.142395$ (861)  \\
  4 & $-176.142513$ (2907) & $-76.080634$ (140) & $-100.048023$ (13) & $-176.142439$ (985)  \\
  5 & $-176.142601$ (3239) & $-76.080686$ (151) & $-100.048023$ (13) & $-176.142533$ (1197) \\
 10 & $-176.142652$ (4150) & $-76.080688$ (184) & $-100.048023$ (13) & $-176.142607$ (1789) \\
\hline
RHF & $-176.085325$ (1)    & $-76.041090$ (1)   & $-100.031644$ (1)  & $-176.085325$ (1) \\
full-CI & $-176.142717$ (27225) & $-76.080693$ (225) & $-100.048023$ (25) & $-176.142717$ (27225) \\
\hline
\end{tabular*}
\end{table*}

\begin{table}[h]
\caption{\label{tab:table_s7} Number of sampled bit strings corresponding to intra-monomer excited and inter-monomer charge-transfer excited Slater determinants.}
\begin{tabular*}{\linewidth}{@{\extracolsep{\fill}}cccc}
\hline
Molecule & $K$ & Intra-monomer & Charge transfer \\
\hline
FH dimer &  1 &  40 & 37 \\
         &  2 &  46 & 51 \\
         &  3 &  54 & 60 \\
         &  4 &  66 & 74 \\
         &  5 &  68 & 77 \\
         & 10 &  87 & 102 \\
FH--H$_2$O &  1 & 117 &  84 \\
           &  2 & 151 & 113 \\
           &  3 & 184 & 160 \\
           &  4 & 210 & 199 \\
           &  5 & 235 & 241 \\
           & 10 & 328 & 409 \\
\hline
\end{tabular*}
\end{table}

\end{document}